\documentclass{bmcart}

\usepackage[utf8]{inputenc} %unicode support
\usepackage[pdftex]{hyperref}
\usepackage{graphicx}
\usepackage{amsthm,amsmath} %for equations
\usepackage{tabularx}
\usepackage{colortbl, color}
\usepackage{xcolor}
\usepackage{booktabs} % for \toprule, etc.
\startlocaldefs
\endlocaldefs

\begin{document}
\begin{frontmatter}

\begin{fmbox}
\dochead{Research}

%%%%%%%%%%%%%%%%%%%%%%%%%%%%%%%%%%%%%%%%%%%%%%
%%                                          %%
%% Enter the title of your article here     %%
%%                                          %%
%%%%%%%%%%%%%%%%%%%%%%%%%%%%%%%%%%%%%%%%%%%%%%

\title{Linguistic neighbourhoods: Explaining cultural borders on Wikipedia through multilingual co-editing activity}

% Explaining cultural borders on Wikipedia through co-editing patterns

% Linguistic neighbourhoods:  Quantifying similarity of interests between cultural communities of Wikipedia editors

%%%%%%%%%%%%%%%%%%%%%%%%%%%%%%%%%%%%%%%%%%%%%%
%%                                          %%
%% Enter the authors here                   %%
%%                                          %%
%% Specify information, if available,       %%
%% in the form:                             %%
%%   <key>={<id1>,<id2>}                    %%
%%   <key>=                                 %%
%% Comment or delete the keys which are     %%
%% not used. Repeat \author command as much %%
%% as required.                             %%
%%                                          %%
%%%%%%%%%%%%%%%%%%%%%%%%%%%%%%%%%%%%%%%%%%%%%%

\author[
   addressref={aff1,aff3},                   % id's of addresses, e.g. {aff1,aff2}
   corref={aff1},                       % id of corresponding address, if any
   %noteref={n1},                        % id's of article notes, if any
   email={anna.samoilenko@gesis.org}  % email address
]{\inits{AS}\fnm{Anna} \snm{Samoilenko}}
\author[
   addressref={aff1},
   email={fariba.karimi@gesis.org}
]{\inits{FK}\fnm{Fariba} \snm{Karimi}}
\author[
   addressref={aff2},
   email={daniel.edler@umu.se}
]{\inits{DE}\fnm{Daniel} \snm{Edler}}
\author[
   addressref={aff3},
   email={kunegis@uni-koblenz.org}
]{\inits{JK}\fnm{Jérôme} \snm{Kunegis}}
\author[
   addressref={aff1,aff3},
   email={markus.strohmaier@gesis.org}
]{\inits{MS}\fnm{Markus} \snm{Strohmaier}}

\address[id=aff1]{%                           % unique id
  \orgname{GESIS -- Leibniz-Institute for the Social Sciences}, % university, etc
  \street{6-8 Unter Sachsenhausen},                     %
  \postcode{50667}                                % post or zip code
  \city{Cologne},                              % city
  \cny{Germany}                                    % country
}

\address[id=aff2]{%                           % unique id
  \orgname{Integrated Science Lab, Department of Physics, Ume\r{a} University}, % university, etc
  \city{Ume\r{a}},                              % city
  \cny{Sweden}                                    % country
}

\address[id=aff3]{%                           % unique id
  \orgname{University of Koblenz-Landau}, % university, etc
  \city{Koblenz},                              % city
  \cny{Germany}                                    % country
}

\end{fmbox}% comment this for two column layout

%%%%%%%%%%%%%%%%%%%%%%%%%%%%%%%%%%%%%%%%%%%%%%
%%                                          %%
%% The Abstract begins here                 %%
%%                                          %%
%% Please refer to the Instructions for     %%
%% authors on http://www.biomedcentral.com  %%
%% and include the section headings         %%
%% accordingly for your article type.       %%
%%                                          %%
%%%%%%%%%%%%%%%%%%%%%%%%%%%%%%%%%%%%%%%%%%%%%%

%%%%%%%%%%%%%%%%%%%%%%%%%%%%%%%%%%%%%   Abstract & keywords %%%%%%%%%%%%%%%%%%%%%%%%%%%%%%%%%%%%%%%%%
\begin{abstractbox}
\begin{abstract}
In this paper, we study the network of global interconnections between language communities, based on shared co-editing interests of Wikipedia editors, and show that although English is discussed as a potential lingua franca of the digital space, its domination disappears in the network of co-editing similarities, and instead local connections come to the forefront. Out of the hypotheses we explored, bilingualism, linguistic similarity of languages, and shared religion provide the best explanations for the similarity of interests between cultural communities. Population attraction and geographical proximity are also significant, but much weaker factors bringing communities together. In addition, we present an approach that allows for extracting significant cultural borders from editing activity of Wikipedia users, and comparing a set of hypotheses about the social mechanisms generating these borders. Our study sheds light on how culture is reflected in the collective process of archiving knowledge on Wikipedia, and demonstrates that cross-lingual interconnections on Wikipedia are not dominated by one powerful language. Our findings also raise some important policy questions for the Wikimedia Foundation.
\end{abstract}

%%%%%%%%%%%%%%%%%%%%%%%%%%%%%%%%%%%%%%%%   Keywords   %%%%%%%%%%%%%%%%%%%%%%%%%%%%%%%%%%%%%%%%%%%%%%%

\end{abstractbox}
\end{frontmatter}

%%%%%%%%%%%%%%%%%%%%%%%%%%%%%%%%%%%%%%%%%%%%%%%%%%%%%%%%%%%%%%%%%%%%%%%%%%%%%%%%%%%%%%%%%%%%%%%%%%%%%%
%%%%%%%%%%%%%%%%%%%%%%%%%%%%%%%%%%%%%%%%%% Article text   %%%%%%%%%%%%%%%%%%%%%%%%%%%%%%%%%%%%%%%%%%%%
%%%%%%%%%%%%%%%%%%%%%%%%%%%%%%%%%%%%%%%%%%%%%%%%%%%%%%%%%%%%%%%%%%%%%%%%%%%%%%%%%%%%%%%%%%%%%%%%%%%%%%

%%%%%%%%%%%%%%%%%%%%%%%%%%%%%%%%%%%%%%%%%% Introduction   %%%%%%%%%%%%%%%%%%%%%%%%%%%%%%%%%%%%%%%%%%%%
\section{Introduction}
%Rapid developments in communication technologies, cheaper transnational travel, and increase in access to education bring nations in contact more frequently than at any previous point in time. Therefore, solving political, environmental and economic challenges of the century requires understanding similarities and differences between communities of different languages and cultures.
%In recent years, multiple approaches have been proposed to quantify similarity between linguistic and cultural points of view, but unfortunately these efforts have not yet crystallised in a holistic, scalable approach for studying general cultural similarities on a global level. To address this challenge, we first need to distinguish collective identities related to culture.

%Documenting accumulated knowledge is one way of building a cultural identity. By writing about people, events and places, members of a language community give them prominence and collectively agree about their importance within this language community \cite{Kramsch}.

Measuring the extent to which cultural communities overlap via the knowledge they preserve can paint a picture of how culturally proximate or diverse they are. Wikipedia, the largest crowd-sourced encyclopedia today, is a platform that documents knowledge from different cultural communities via different language editions. The collective traces left by editors of Wikipedia can be utilized to identify cultural communities that are most similar with regard to the knowledge they document. Certainly, co-editing similarities among language communities of Wikipedia editors are just a particular dimension of culture and are not representative of cultural similarities among the communities in general. Yet, Wikipedia plays a critical role in today's information gathering and diffusion processes and Wikipedians constitute an important cultural subset of educated and technology-savvy elites who often drive the cultural, political, and economic processes \cite{Ronen}. In this paper, we tap into the traces left by editors of Wikipedia to gain new insights into how language communities on Wikipedia relate to each other via common co-editing interests.

\textbf{Problem.}
We are thus interested in seeking answers to the following overarching research question: What are common editing interests between language communities on Wikipedia, and how can they be explained? In addition, we also aim to establish a computational method which would allow measuring culture-related similarities based on the topics the editors document in Wikipedia.

We assume that collective interest of a language-speaking community is reflected through the aggregation of articles documented in the corresponding language edition of Wikipedia. These articles are an approximation of the topics which are culturally relevant to that language community, though by no means are representative of the entire underlying cultural community. We define \textit{cultural similarity} as a significant interest of communities in editing articles about the same topics; in other words, language communities are similar when they significantly agree regarding the topics they choose to edit.

\textbf{Methods.}
Our approach consists of several steps. We first use statistical filtering to identify language pairs which show consistent interest in articles on the same topics. Based on this dyadic information, we create a network of interest similarity where nodes are languages and links are weighted as the strength of shared interest. We cluster the network and inspect it visually to inform the generation of hypotheses about the mechanisms that contribute to cultural similarity. Finally, we express these hypotheses as transition probability matrices, and test their plausibility using two statistical inference techniques -- HypTrails~\cite{hyptrails} and MRQAP \cite{krackardt1987qap} (Multiple Regression Quadratic Assignment Procedure). Using both Bayesian and frequentist approaches, we obtain similar results, which suggests that our findings are robust against the chosen statistical measure.

\textbf{Contribution and findings.}
Our main contribution is empirical. We expand the literature on culture-related research by (a)~presenting a large-scale network of interest similarities between 110 language communities, (b)~showing that the set of languages covering a concept of Wikipedia is not a random choice, and (c)~by statistically demonstrating that similarity in concept sets between Wikipedia editions is influenced by multiple factors, including bilinguality, proximity of these languages, shared religion, and population attraction. We also combine multiple techniques from network theory, Bayesian and frequentist statistics in a novel way, and present a generalisable approach to quantify and explain culture-related similarity based on editing activity of Wikipedia editors. 

We find that the topics that each language edition documents are not selected randomly, however small the underlying community of editors. We test several hypotheses about the underlying processes that might explain the observed nonrandomness, and find that bilingualism, linguistic similarity of languages, and shared religion provide the best explanations for the similarity of interests between cultural communities. Population attraction and geographical proximity are also significant, but much weaker factors bringing communities together.

The remainder of the paper is structured as follows. In Related Literature (Section \ref{literature}) we will give a brief overview of work on how cultural differences find reflection in multilingual online platforms, as well as on how Wikipedia has been used to compare cultural and linguistic points of view, and cultural biases involved in knowledge production. In the Data, Section \ref{data} we will describe in detail the process of data sampling and collection. Sections \ref{4} and \ref{5} will focus on identifying and explaining co-editing interests, give a technical overview of the quantitative methods, and report the results. We will offer our reflection upon the findings in the Discussion (Section \ref{discussion}), and Conclusions and Implications (Section \ref{conclusions}).

%%%%%%%%%%%%%%%%%%%%%%%%%%%%%%%%%%%%%   Background   %%%%%%%%%%%%%%%%%%%%%%%%%%%%%%%%%%%%%%%%%%
\section{Related literature} \label{literature}
Definition of culture and its borders is a long-debated and still unresolved issue in Anthropology and Social Sciences; a 1951 review of the works on the issue already contained  close to 300 definitions of culture \cite{Kroeber}. Cultural communities have fuzzy boundaries: several distinct cultures might co-exist in one state, or alternatively, reach beyond and across continents. This is especially true for multilingual countries or those with colonial past. While there are many non-verbal expressions of material culture, language is an important bearer of culture -- its meanings have to be learnt socially and represent the way of life as seen by a particular community \cite{Bloomfield, Hoijer, Silvia, Voegelin}. Language-speaking communities form distinct and unique cultures around themselves \cite{Bucholtz, Geertz-1973}, and overlap of interests between these communities might signify cultural proximity between them. Language is central to culture for several reasons:  it reflects the collective agreement of a language community to view the world in a certain way, and helps a community to perpetuate its culture, develop its identity, and archive accumulated knowledge \cite{Kramsch}. It is the latter feature of collective knowledge selection and archiving that this paper focuses on.

\textbf{Wikipedia as a lens for studying cultural repertoires of language communities. }
The online encyclopedia Wikipedia is a prominent example of collective knowledge accumulation, and it is becoming one of the most interesting and convenient sources for academics to study cultural and historical processes \cite{Schich-15}. Wikipedia is one of the most linguistically diverse projects online, with a constant base of editors contributing in almost 300 languages \cite{WP:list}, ranging from almost 5 million in the largest edition (English) to just 89 in Cree, the smallest one \cite{WP:list}. This makes it accessible to more than 5 billion people, or 75\% of the world's population \cite{Petzold}. There is no central authority that dictates which topics must be covered, and every editor is free to select their own, as long as they are consistent with the notability guidelines \cite{WP:notability}. All language editions have their own notability guidelines and are edited independently from each other, although an editor can also co-edit several editions in parallel. Large language editions like English are not supersets of smaller ones, and each edition contains unique concepts which are not covered by others. For example, concept overlap between the two largest editions, English and German, is only 51\% \cite{Hecht:2010content}. Opposite to the common misconception, even when articles on the same concept exist in different language editions, they are not translated replicas of each other, but instead reveal consistent cultural biases \cite{Hecht:2009bias, Callahan-2011} and introduce various linguistic viewpoints \cite{Bao-2012, Massa-2012, Massa-2011}.

These differences in number, selection, and content of articles across languages are not accidental, but relate to the cultural differences between the underlying language communities. Contributing to Wikipedia means more than writing encyclopedic content: it allows communities to store cultural memories of events \cite{Keegan-2011, Keegan-newswork-2013, Pentzold-12}, document their point of view \cite{Massa-2012, Massa-2011}, and give prominence to people \cite{Samoilenko-2014}. This collective sifting of culturally-relevant knowledge is such an important social process that conflicts and edit wars frequently emerge before reaching consensus \cite{Yasseri-book}. Finally, the language communities not yet represented on Wikipedia seek the inclusion as an opportunity to establish and promote their language and culture in the digital realm \cite{Kornai-2013}. There are currently 160 open requests for new Wikipedia language editions in the Wikimedia Incubator \cite{WP-incubator}. Wikipedia is rich in cultural material, and all data are recorded and openly available, which makes the encyclopedia an attractive object for research on culturally-mediated behaviour.

\textbf{Quantifying cultural similarity. }
Multiple numerical measures have been proposed to assess the degree of cultural similarity, although many of them suffer from practical scalability issues or focus on a narrow aspect of culture. The most often cited measure is known as Hofstede's dimensions of culture, which delineates cultures by national borders \cite{Hofstede-1980}. Evidence of  national cultural differences has been found in the style of collaborative authoring of Wikipedia articles \cite{Pfeil-2006, Hara-2010}. West \cite{West-2004} quantifies cultural distance through linguistic distance between languages. Several studies delineated cultures by language, and focused on Wikipedia data. In particular, Laufer and colleagues~\cite{Laufer-2014} developed measures of cultural similarity, understanding, and affinity through comparing how food cultures are described by self- and foreign communities. Eom et al.~\cite{Eom-2015} applied ranking algorithms to biographical articles and obtained a network of cultural agreement on what historical figures are viewed as important, which includes 24 language points of view. Finally, the value of Wikipedia for such anthropological questions as assessing cultural chauvinism or differences in historical world view between cultures has been discussed in \cite{Gloor}. Cultural differences have also been found in other modalities of online communication and collaboration, on such multilingual platforms as Facebook \cite{Barnett-FB}, Twitter \cite{Eleta-4, Garcia-6, Mocanu-11}, and YouTube \cite{Platt-14}.

Although previous research has advanced scientific understanding of cultural similarity, attempts to quantify it, for practical reasons, were mostly limited to comparing a small number of cultures along a selected topical dimension. The literature shows a need to establish a scalable approach to quantifying cultural similarity which allows comparing multiple permutations of language dyads and obtaining a bird's-eye view on global intercultural relationships.

%%%%%%%%%%%%%%%%%%%%%%%%%%%%%%%%%%%%%   Data and Methods   %%%%%%%%%%%%%%%%%%%%%%%%%%%%%%%%%%%%%%%%%%
\section{Data} \label{data}
%%%%%%%%%%%%%%%%%%%%%%%%%%%%%%%%%%%%%%%%%%%   Data   %%%%%%%%%%%%%%%%%%%%%%%%%%%%%%%%%%%%%%%%%%%%%%%%%
There are almost 300 language editions of the encyclopedia, which vary greatly in size. This makes sampling a nontrivial decision: on the one hand, many editions are rather small, and sampling from them would not provide data sufficient for statistical analysis. On the other hand, downloading full data on every language edition over a long period of time would be computationally expensive. As a compromise, we focused the analysis on a sample of 126 largest editions which contained more than 10,000 article pages, as of July 2014 ~\cite{WP:list}.

\textbf{Sampling procedure. }
To account for variations in editions' age, number of active contributors, and growth rates, we selected the time frame such that (1) to ensure a sufficient amount of editions existed in the beginning of the observation; (2) to allow enough time for each edition to accumulate concepts. We traced back each edition to its first registered article page, and found out that 110 out of 126 largest editions had been created before 01.01.2005. We excluded 11 editions which appeared later (min, vo, be, new, pms, pnb, bpy, arz, mzn, sah, vec) and those whose language codes could not be mapped to the ISO 639-1 standard (be-x-old, zh-yue,bat-smg, map-bms, zh-min-nan). These remaining 110 editions became the focus of our subsequent analysis which covers the period of 9 years between 01.01.2005 and 31.12.2013.

We sampled from each edition separately, collecting IDs of all article pages created between 2005 and 2013 (excluding other types of pages, redirects, and pages created by bots). For each ID we also collected the entire editing history in all linked language editions. Thus, each ID corresponds to a concept (the topic of the article regardless of the language), and all interlinked language editions represent various linguistic points of view on the concept. After removing duplicates, our dataset includes 3,066,736 unique concepts and a total of 1,360,647,795 article pages in different languages. The data were collected between 20.12.2015 and 25.01.2016 from Wikimedia servers directly, using the access provided by Wikimedia Tool Labs~\cite{toolserver}.

One algorithmic limitation of our approach is the fact that we rely on Wikipedia's interlanguage link graph to identify articles on the same concepts in different language editions. This approach has some known issues with the lack of triadic closure and dyadic reciprocity \cite{Bao-2012}. To ensure that the maximal set of interlanguage links related to a concept is retrieved, we collect all articles with their interlanguage links from each edition separately, removing duplicates afterwards. Thus, all existing interlanguage links are extracted.

%%%%%%%%%%%%%%%%%%%%%%%%%%%%%%%%%%%%%%%%   Identifying co-editing patterns   %%%%%%%%%%%%%%%%%%%%%%%%%%%%%%%%%%%%%%
\section{Extraction of co-editing patterns} \label{4}
In this section, we describe the procedure of extracting cultural similarities from co-editing activity in Wikipedia, and present the network of significant shared interests between 110 language communities. The section begins with summarising our pre-analysis check of whether the language-concept overlap in Wikipedia is random.

\subsection{Testing for non-randomness of co-editing patterns}
Theoretically, each concept covered in Wikipedia could exist in all 288 language editions of the encyclopedia. This is possible because Wikipedia does not censor topic inclusion depending on the language of edition, and anyone is free to contribute an article on any topic of significance. However in practice, such complete coverage is very rare, and concepts are covered in a limited set of language editions. Is this set of languages random? To answer this question, we analyse matrices of language co-occurrences based on a 6.5\% random sample of the data (200,748 concepts).

We construct the matrix of empirical co-occurrences $C_{ij}$, based on the probability of languages ${i,j}$ to have an article on the same concept. 
We also construct a synthetic dataset where we preserve the distribution of languages and the number of concepts, $N = 200{,}748$, but allow languages to co-occur at random. We use the resulting data to produce the matrix of random co-occurrences $C^{\mathrm{rand}}_{ij}$, and compare it to the matrix of co-occurrences $C_{ij}$. Our null model corresponds to belief that in Wikipedia each concept has equal chances to be covered by any language, with larger editions sharing concepts more frequently purely because of their size. Comparing the two matrices (Figure~\ref{fig:SI} in the Appendix) allows us to get a preliminary intuition of the extent to which co-editing patterns are non-random.

We establish that language dyads do not edit articles about the same concept (co-occur) by chance. Large editions share concepts more frequently than expected: although in the data \emph{EN-DE} and \emph{EN-FR} overlap in 45\% of cases, only 15\% is expected by the null model. To little surprise, the amount of overlap between editions in the data decreases with the size of the editions. One notable exception is the Japanese edition which, despite being among the ten largest Wikipedias, co-occurs with other top editions noticeably less frequently. Similarly, the Uzbek edition, being among the ten smallest in the dataset, shows high concept overlap with large editions. By simply plotting frequencies of co-occurrences, we do not observe any local blocks or clusters, neither among large nor small editions (see Fig.~A1).

These overlap differences are statistically significant, and the null model explains only 1,386 out of 11,990 language pairs (11\% of observed data, 95\% confidence level). Such low explained variation suggests that concept overlap is not random and cannot be explained only by edition sizes. Instead, there are non-random, possibly cultural processes, that influence which languages cover which concepts on Wikipedia. Having evidence that the data contain a signal, we continue our investigation by performing network analysis.

%%%%%%%%%%%%%%%%%%%%%%%%%%%%%%%%%%%%%%%%   Methods: Filtering   %%%%%%%%%%%%%%%%%%%%%%%%%%%%%%%%%%%%%%
\subsection{Inferring the network of shared interest}
\begin{figure}
    \centering
    \includegraphics[width=0.97\textwidth]{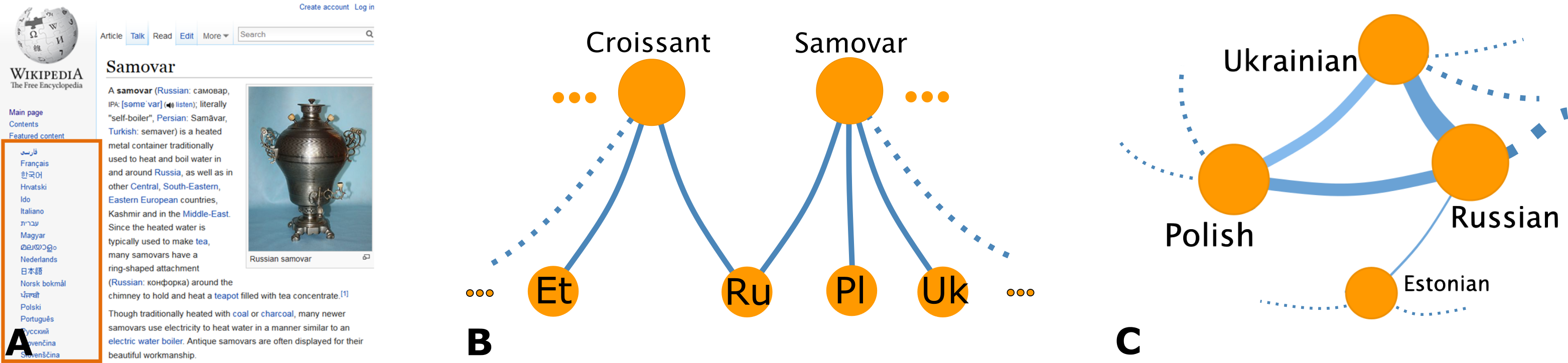}
    %% \begin{tabular}{p{0.28\textwidth} p{0.28\textwidth} p{0.28\textwidth}}
    %%   \includegraphics[width=0.27\textwidth]{img/fig1a} &
    %%   \includegraphics[width=0.27\textwidth]{img/fig1b} &
    %%   \includegraphics[width=0.27\textwidth]{img/fig1c} \\
    %%   (a) Retrieving edits to a concept in all linked language editions &
    %%   (b) Fragment of the unfiltered bipartite co-editing network &
    %%   (c) Fragment of the filtered co-editing network with significant links
    %% \end{tabular}
    %% \subfigure[Retrieving edits to a concept in all linked language editions.]{
    %%    \includegraphics[width=0.27\textwidth]{img/fig1a}
    %%   \label{fig1a}
	%% }\hfill
	%% \hspace{2mm}
	%% \subfigure[Fragment of the unfiltered bipartite co-editing network]{
	%%     \raisebox{0.5\height}{
	%% 	\includegraphics[width=0.27\textwidth]{img/fig1b}}
	%% 	\label{fig1b}
	%% }\hfill
	%% \hspace{2mm}
	%% \subfigure[Fragment of the filtered co-editing network with significant links]{
	%%     \raisebox{0.2\height}{
	%% 	\includegraphics[width=0.27\textwidth]{img/fig1c}}
	%% 	\label{fig1c}
	%% }\hfill
	\caption{\csentence{Illustration of the $z$-score-based filtering method.} The method requires three steps: (a)~to retrieve all edits to each concept in all linked language editions; (b)~to compare the empirical and expected probabilities of each language pair to co-edit a concept; and  (c)~to create a filtered network of languages with significant shared interests. In the final network, `heavier' links signify stronger co-editing similarity between the nodes.}
    \label{fig1}
\end{figure}

We look for the languages that are consistently interested in editing articles on the same topics by comparing the differences between observed and expected co-editing activity on each concept. We give a $z$-score to every language pair, and compare it to the threshold of significance to filter out insignificant pairs. This logic is demonstrated in Fig.~\ref{fig1}. The result is a weighted undirected network of languages, where languages are connected based on shared information interest.

% Fig 1

We first compute the empirical weight $w_{ij}^{c}$ of a link between languages $i$, $j$ which co-edit a concept $c$:
\begin{equation}
   w_{ij}^{c} = k_{i}^{c} k_{j}^{c}.
   \label{eq:empirical_weight}
\end{equation}

Here, $k_{i}^{c}$ is the number of edits to the concept $c$ in the language edition $i$, which we use as a proxy to the amount of editing work invested in the concept. This is done across all concepts and language permutations. To determine which links are statistically significant, and which exist purely by chance or due to size effects, we construct a null model where we assume that links between languages $i$ and $j$ are random.

Let the total editing probability of a language be $p_{i} = \frac 1 M \sum_{c}k_{i}^c$, where $M$ is the total number of edits for all concepts and language editions. Then the expected probability $\mathrm{E}[w_{ij}^{c}]$ that languages $i$ and $j$ co-edit the same concept $c$ is:\begin{equation}
  \mathrm E[w_{ij}^{c}] = n_{c} (n_{c} - 1) p_{i} p_{j},
 \label{eq:expected_weight}
  \end{equation}
where $n_{c}$ is the total number of edits to a concept from all language editions. To compare the difference between observed and expected link weights, we compute a $z$-score $z_{ij}^{c}$ for each concept and pair of languages ${i,j}$, defined as
\begin{equation}
  z_{ij}^{c} = \frac{w_{ij}^{c} - \mathrm E[w_{ij}^{c}]}{\sigma_{ij}^{c}},
  \label{eq:z_scores}
\end{equation}
where $\sigma_{ij}^{c}$ is the standard deviation of the expected link weight \cite{Karimi}.

Finally, to find the cumulative $z$-score for a pair of languages ${i,j}$, we sum their $z$-scores over all concepts\begin{equation}
z_{ij} = \sum_{c}z_{ij}^{c}.
 \label{eq:final_z}
  \end{equation}
The relationship between $i$ and $j$ is significant if the cumulative probability of their total $z$-score, $z_{ij}$ in the right tail falls beyond the $p$-value $p = 1 - 0.05 / N$, where $N$ is the total number of languages. We use the Bonferroni correction \cite{dunn1961multiple} to account for the multiple comparisons and size effects in the data. This corresponds to a $z$-score of 3.32. Since $z$-scores are sums across many independent variables, their distribution can be approximated by the normal distribution, and the threshold for link significance in the right tail is $t = 3.32 \sqrt{L}$, where $L = 3{,}066{,}736$ is the number of concepts. We create a link between a pair of languages ${i,j}$ if the observed $z$-score, $z_{ij}$, is above the threshold $t$~\cite{Karimi}.

We use the resulting $z$-scores to build a network of shared topical interests, where the edges are weighted by the similarity of interest, quantifies via $z$-scores. In summary, this approach allows for discovering significant language pairs of shared interest, accounting for editions of different sizes, and avoiding over-representing the large editions \cite{Karimi}.
%Importantly, it brings out the connections between smaller editions.

Other methods exist to extract significant weights in graphs. For example, \cite{Tumminello} used the hypergeometric distribution for finding the expected link weights for bipartite networks and measured the global $p$-value. Serrano et al.~\cite{Serrano} used a disparity filtering method to infer significant weights in networks. Similar to our work, \cite{Dianati} proposed pair-wise connection probability by the configuration model and used the $p$-value to measure statistical significance of the links.
%The advantage of our model is that it accounts for variations in the number of edits to an article, and the editing activity level of each language edition.

The network consists of 110 nodes (language editions) and 11,986 undirected edges, and is a complete graph. This means that most languages show at least some similarity in the concepts they edit, however the strength of similarity differs highly across language pairs. The distribution of edge weights is highly skewed with the lowest $z$-score between Korean and Buginese  and the highest $z-$-score between Javanese and Indonesian.

\subsection{Clustering the network of significant shared interests}

We use the Infomap algorithm~\cite{rosvall} to identify language communities that are most similar in their interests. We release a random walker on the network, and allow it to travel across links proportional to their weights. By measuring how long the random walker spends in each part of the network, we are able to identify clusters of languages with strong internal connections \cite{rosvall}. Additionally, we compare these results with the Louvain clustering algorithm~\cite{louvain} and establish that both methods show high agreement.

Our cluster analysis suggests that no language community is completely separated from other communities, and in fact, there are significant topics of common interest between almost any two language pairs. We reveal 21 clusters of two and more languages, plus 9 languages that are identified as separate clusters (see SI for full information on the clusters). Notably, English forms a self-cluster, and this independent standing means little interest similarity between English and other languages. This is an interesting finding in the light of the recent discussions on whether English is becoming a global language and the most suitable \textit{lingua franca} for cross-national communication \cite{crystal:global}.

% Fig 2
\begin{figure*}
  \centering
  \includegraphics[width=0.97\textwidth]{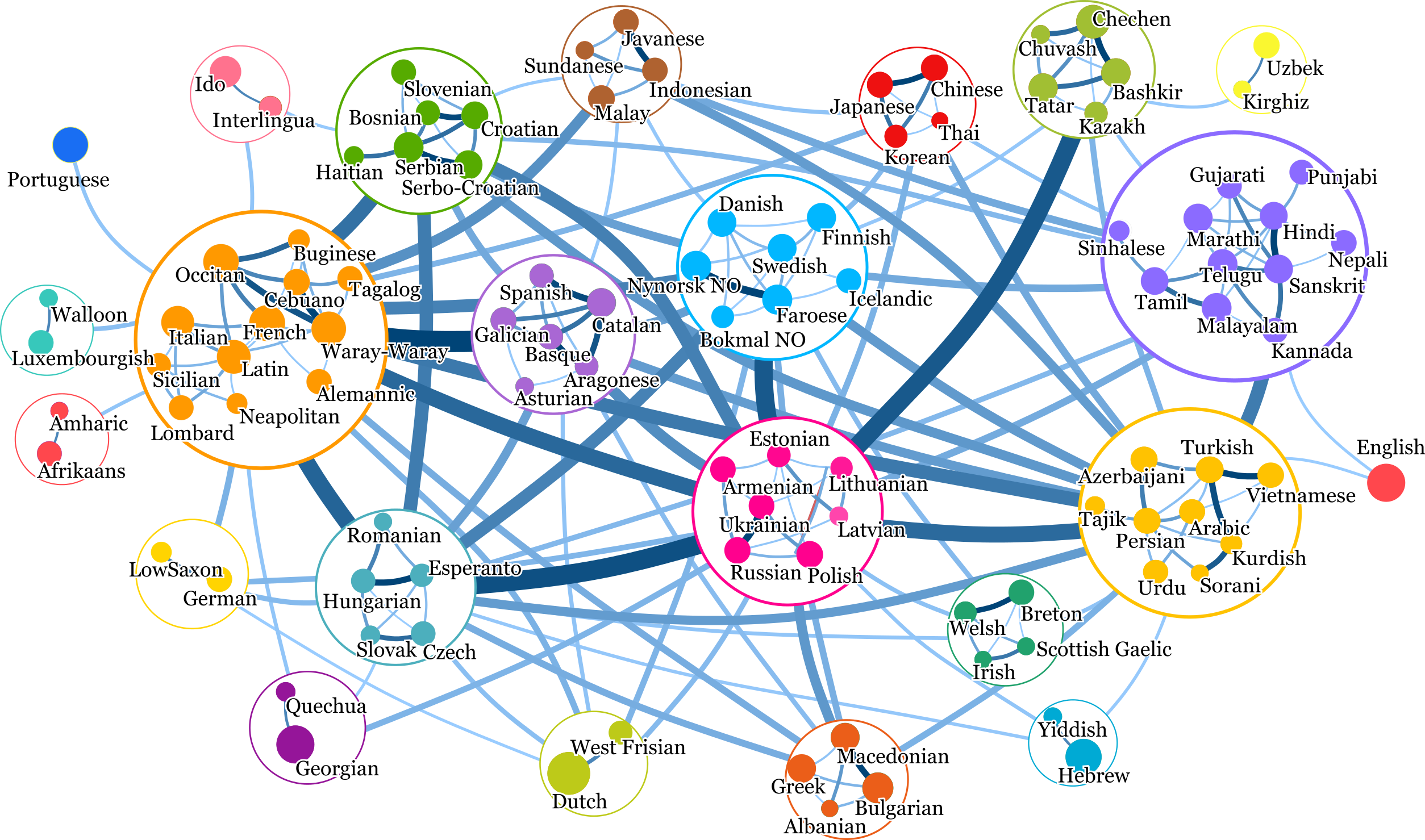}
  \caption{\csentence{The network of significant Wikipedia co-editing ties between language pairs.} Nodes are coloured according to the clusters found by the Infomap algorithm \cite{Edler-2013}, and link weights within clusters represents the positive deviation of $z$-scores from the threshold of randomness; links are significant at the 99\% level. For visualisation purposes we display only 23 clusters and the strongest inter-cluster links in the network. The inter-cluster links show the aggregated z-scores between all nodes of a pair of clusters. The network suggests that local factors such as shared language, linguistic similarity of languages, shared religion, and geographical proximity play a role in interest similarity of language communities. Notably, English forms a separate cluster, which suggest little interest similarity between English speakers and other communities.
\label{fig:network}
}
\end{figure*}

The resulting network is visualised in Fig.~\ref{fig:network}. The links within clusters are weighted according to the amount of positive deviation of $z$-score per language pair from the threshold of randomness. Stronger weights indicate higher similarity. The links are significant at the 99\% level. The inter-cluster links should be interpreted with care in the context of this study, as they are weighted according to the aggregated strength of connection between all nodes of both clusters. The network is undirected since it depicts mutual topical interest of both language communities, which is inherently bidirectional. For visualisation purposes, we display only the strongest inter-cluster links and 23 language clusters. Cluster membership information is detailed in Table~\ref{SI_table} (in the Appendix).

\textbf{Cluster interpretation. }
Visual inspection of language clusters suggests a number of hypotheses which might explain such network configuration. For example, (1) geographical proximity might explain the Swedish-Norwegian-Danish-Faroese-Finnish-Icelandic cluster (light blue), since those are the languages mostly spoken in Scandinavian countries. Other groups of languages form around (2) a local \textit{lingua franca}, which is often an official language of a multilingual country, and include other regional languages which are spoken as second- and even third language within the local community. This way, Indonesian and Malay form a cluster with Javanese and Sundanese (brown), which are the two largest regional languages of Indonesia. Similarly, one of the largest clusters in the network (purple) consists of 11 languages native to India, where cases of multilingualism are especially common, since one might need to use different languages for contacts with the state government, with the local community, and at home \cite{crystal:global}. Another interesting example is the cluster of languages primarily spoken in the Middle Eastern countries (yellow), which apart from geographical proximity are closely intertwined due to (3) a shared religious tradition. Finally, some clusters illustrate (4) the recent changes in sociopolitical situation, which can also be partially traced through bilingualism. Following the civil war of the 1990s in former Yugoslavia, its former official Serbo-Croatian language is now replaced by three separate languages: Serbian, Croatian, and Bosnian (green cluster). Notably, there is still a separate Serbo-Croatian Wikipedia edition. To give another example, Russian held a privileged position in the former Soviet Union, being the language of the ideology and a priority language to learn at school \cite{crystal:global}. Even twenty years after the dissolution of the Soviet Union, Russian remains an important language of exchange between the post-Soviet countries. Similarity of interests between speakers of Russian and the languages spoken in nearby countries, as seen in the magenta cluster, comes as little surprise.

We use this anecdotal interpretation of the clusters to inform our hypotheses about the mechanisms that affect the formation of co-editing similarities. In the next section we will build on these initial interpretations and formulate them as quantifiable hypotheses. To evaluate the validity of the hypotheses, we will compare their plausibility against one another using statistical inference approach.

%%%%%%%%%%%%%%%%%%%%%%%%%%%%   Explaining the clusters  %%%%%%%%%%%%%%%%%%%%%%%%%%%%%%%%%%%%%%%%%%%%%%%%%
\section{Explanation of co-editing patterns} \label{5}
In this section we show how the network of significant shared interests could be used to inform hypothesis formulation. We compare the plausibility of hypotheses using two statistical approaches. First, we use Bayesian approach and visually compare the strengths of hypotheses. Then we apply frequentist approach to report the explanatory power of different models. We begin by outlining the necessary methodology and continue with reporting the results.

%%%%%%%%%%%%%%%%%%%%%%%%%%%%%%%%%%%%%%%%   Methods: Hyptrails   %%%%%%%%%%%%%%%%%%%%%%%%%%%%%%%%%%%%%%
\subsection{Hypothesis formulation}
\begin{figure}
    \centering
    \includegraphics[width=0.97\textwidth]{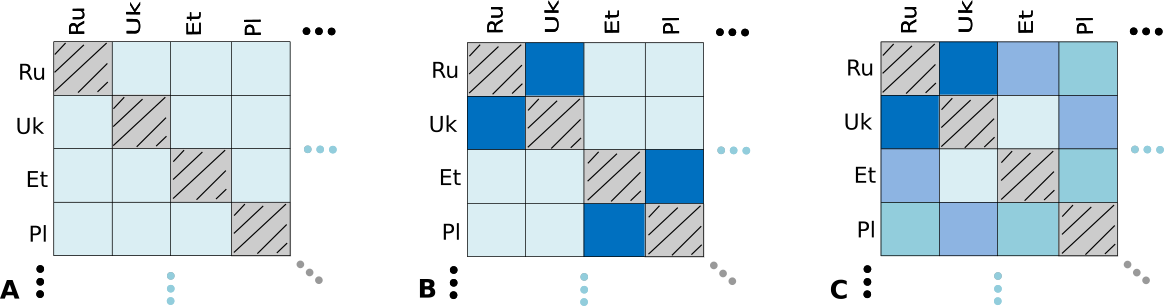}
    %% \begin{tabular}{p{0.28\textwidth} p{0.28\textwidth} p{0.28\textwidth}}
    %%   \includegraphics[width=0.27\textwidth]{img/fig2a} &
    %%   \includegraphics[width=0.27\textwidth]{img/fig2b} &
    %%   \includegraphics[width=0.27\textwidth]{img/fig2c} \\
    %%   (a) Uniform hypothesis  &
    %%   (b) Shared religion hypothesis &
    %%   (c) Geographical proximity hypothesis
    %% \end{tabular}
	%% \subfigure[Uniform hypothesis]{
	%% 	\includegraphics[width=0.27\textwidth]{img/fig2a}
	%% 	\label{fig2a}
	%% }\hfill
    %% 	\hspace{2mm}
	%% \subfigure[Shared religion hypothesis]{
	%% 	\includegraphics[width=0.27\textwidth]{img/fig2b}
	%% 	\label{fig2b}
	%% }\hfill
	%% \hspace{2mm}
	%% \subfigure[Geographical proximity hypothesis]{
	%% 	\includegraphics[width=0.27\textwidth]{img/fig2c}
	%% 	\label{fig2c}
	%% }\hfill
	\caption{\csentence{A toy example of expressing a hypothesis through a transition probability matrix.} The matrices are symmetrical. The diagonal is empty since the data do not allow self-loops. According to each hypothesis, the cells with more likely transitions are coloured in darker shades of blue. In (a) Uniform hypothesis -- all transitions are equally possible, i.e. the editions are covering random topics. In (b) Shared religion hypothesis -- the dyads Russian-Ukrainian and Polish-Estonian are given more belief on the basis of shared religion. Finally, in (c) Geographical proximity hypothesis -- the shorter the distance between languages, the stronger belief in the transition.
	}
    \label{fig2}
\end{figure}

We convert our initial interpretation of the network clusters into quantifiable hypotheses, which we express through transition probability matrices illustrated in Fig.~\ref{fig2}. The hypotheses aim to explain the link weights in the network of co-editing similarities, which correspond to the obtained $z$-scores. The transition probability matrices are square with dimensions $N = 110$, corresponding to the number of language editions studied. The diagonal is empty, since self-loops are not allowed. The formulas, the definitions, and data sources for hypotheses formulation are summarised for reference in Table \ref{hypotheses_table}. Below we give more extended explanations on the process of hypotheses construction.
\begin{itemize}
    \item{\textbf{H0: Uniform}}
	
All language co-occurrences are possible with the same probability. A concept can be randomly covered by any language edition. The transition probability $t_{ij}$ for all permutations of languages $i$ and $j$ is \begin{equation*}t_{ij} = 1.\end{equation*} 
    
    \item{\textbf{H1: Shared language family}}

We retrieve the whole family tree profile of each language and count the number of branches overlapping between each language dyad. For example,
    \begin{itemize}
    	\item{Arabic: Afro-Asiatic; Semitic; Central Semitic; Arabic languages; Arabic}
    	\item{Hebrew: Afro-Asiatic; Semitic; Central Semitic; Northwest Semitic; Canaanite; Hebrew}
    \end{itemize}

Arabic and Hebrew share three levels of language tree hierarchy (Afro-Asiatic; Semitic; Central Semitic) and thus will have the transition score of 3 in the hypothesis table. 
If $f_i$ is the set of branches describing the full language family profile of language $i$, the transition probability $t_{ij}$ corresponds to the count of shared branches in the family tree of languages $i$ and $j$, and is computed as \begin{equation*}t_{ij} = |f_i \cup f_j|.\end{equation*}

    \item{\textbf{H2: Bilingual population within a country}}

To formalise other hypotheses, we needed to map languages to countries where they are spoken. We list all countries where a pair of languages are co-spoken; for each country we compute the probability of a person to speak both languages. The hypothesis table contains the average probability of a person to speak both languages computed across all countries where both languages are spoken by more than 0.1\% of the population. The transition probability is described by \begin{equation*} t_{ij} = \frac{1}{N_{ij}} \sum_A p(i)_A p(j)_A,\end{equation*} where $p(i)_A$, $p(j)_A$ are proportions of speakers of languages $i$, $j$ in a country $A$, $N_{ij}$ is the number of countries where $i$,$j$ are co-spoken. The more bilinguals speaking $i$ and $j$ live in the same country, the higher the transition belief.
  
    \item{\textbf{H3: Geographical proximity of language speakers}}

We assign each country with its primary language (the language that the majority of its population speaks) and compute the average distance between all permutations of countries  where language $i$ or $j$ are spoken. All inter-country distances are scaled between 0 and 1. Thus, \begin{equation*} t_{ij} = \frac{1}{N_{ij}} \sum_{A,B}\frac{ d_{\mathrm{min}} }{ d_{AB}},\end{equation*} where $N_{ij}$ is the number of country permutations where $i$ or $j$ are spoken as primary language,  $d_{AB}$ is Euclidean distance between each pair of countries, and $d_{\mathrm{min}}$ is the smallest distance between countries in the dataset. The smaller the distance between speakers of $i$ and $j$ living in separate countries, the higher the chances for languages $i,j$ to cover the same concept.
   
    \item{\textbf{H4: Gravity law -- demographic force attracting language communities}}

Like in the previous example, we allow one (primary) language per country and consider all country permutations where languages $i$ or $j$ are spoken. Demographic attraction is strongest between large population of speakers who live in separate counties which are located closely. Consider the example of France and Germany, where large numbers of French and German speakers correspondingly, live at close distance. We compute average demographic attraction between all permutations of country pairs. We define \begin{equation*} t_{ij} = \frac{1}{N_{ij}}  \sum_{A,B} \frac{m_{A,i}  m_{B,j}}{d_{AB}^2},\end{equation*} where $m_{A,i}$, number of speakers of the primary language $i$ in a country $A$, $d_{AB}$ is Euclidean distance between each pair of counties (in kilometers), $N_{ij}$ is the number of country pairs where $i$ or $j$ are spoken as primary language. The larger the language-speaking population and the smaller the distance between the countries $A,B$, the more the attraction between $i$ and $j$. 

    \item{\textbf{H5: Shared primary religion}}

For each country we identified its primary language and its most widespread religion (Christian, Muslim, Hindu, Buddhist, Folk, other or unaffiliated). The religion we assign to a language is the most common religion in the list of countries where the language is spoken as primary. For a language pair, if they share the religion, we add 1 to the hypothesis matrix, and 0 otherwise. Thus the linguistic communities which profess the same religion will show consistent interest in the same topics.

\end{itemize}

% Table 1
\begin{table}
	\centering
	\caption{
          % \csentence remove because it is only meant for figures, not for tables.
          Formalisation of hypotheses to explain the probability of language dyads to co-edit a Wikipedia article about the same concept.
          The hypotheses aim to explain the values of link weights ($z$-scores) in the network of co-editing similarity (see Fig.\ref{fig:network} for illustrative purposes). The transition probability matrices are square with dimensions $N$ = 110, corresponding to the number of language editions studied. The diagonal is empty, since self-loops are not allowed. The value $t_{ij}$ expresses the hypothesised probability of Wikipedia language editions $i$ and $j$ to cover the same concept. After construction of the hypotheses matrices, the matrices undergo Laplacian smoothing of weight 1 (for HypTrails hypotheses testing only), and are further normalised row-wise. The precess is illustrated in Fig.\ref{fig2}. The results of hypothesis testing are represented in Fig.\ref{fig:hyptrails} for the HypTrails approach, and in Fig.\ref{table:mrqap} for the MRQAP approach, and are discussed in sections \ref{Hyptrails} and \ref{mrqap} correspondingly. 
        }
	\label{hypotheses_table}
\makebox[\textwidth]{
\scalebox{0.93}{
\begin{tabular}{m{4cm}|m{3.3cm}|m{3.3cm}|m{3.3cm}}
\toprule
\textbf{Hypothesis and Formalisation} & \textbf{Notation} & \textbf{Description} & \textbf{Data Source} \\
\midrule
H0: Uniform hypothesis \begin{equation*} t_{ij} = 1 \end{equation*} & -- &
All co-occurrences are equally probable, i.e. every edition $i$ covers the same concept as edition $j$ with a constant probability. &  --
\\
\midrule
H1:  Shared language family
\begin{equation*}t_{ij} = |f_i \cup f_j|\end{equation*}
& $f_i$ is the set of branches describing the full language family profile of language $i$, $t_{ij}$ is the count of shared
  branches in the family tree of $i$ and $j$.
& Language communities of linguistically related languages will show more co-editing similarity.
& The data on language family classification was taken from English Wikipedia infoboxes of articles on each of 110 languages, such as `Hebrew language'.
\\
\midrule
H2: Bilingual population within a country \begin{equation*} t_{ij} = \frac{1}{N_{ij}} \sum_A p(i)_A p(j)_A\end{equation*} &
$p(i)_A$, $p(j)_A$ are proportions of speakers of $i$, $j$ in a country $A$, $N_{ij}$ is the number of countries where $i$,$j$ are co-spoken.     &
Multilingual editors belong to multiple cultural communities and might serve as bridges between them. The more bilinguals speaking $i$ and $j$ live in the same country, the higher the transition belief.
& Territory--language information was downloaded from \cite{data:lang}, and is based on the data from the World Bank, Ethnologue, FactBook, and other sources, including per-country census data.
\\
\midrule
H3: Geographical proximity of languages \begin{equation*} t_{ij} = \frac{1}{N_{ij}} \sum_{A,B}\frac{ d_{\mathrm{min}} }{ d_{AB}}\end{equation*} &
$N_{ij}$ is the number of country permutations where $i$ or $j$ are spoken as primary language,  $d_{AB}$ is Euclidean distance between each pair of countries, and $d_{\mathrm{min}}$ is the smallest distance between countries in the dataset. &
The smaller the distance between speakers of $i$ and $j$ living in separate countries, the higher the chances for languages $i,j$ to cover the same concept. We consider one (primary) language per country.
& Distance between countries is computed as Euclidean distance in kilometers between country capitals \cite{CIA_factbook}.
\\
\midrule
H4: Gravity law -- demographic force attracting language communities
\begin{equation*} t_{ij} = \frac{1}{N_{ij}}  \sum_{A,B} \frac{m_{A,i}  m_{B,j}}{d_{AB}^2}
\end{equation*}
&
$m_{A,i}$, number of speakers of the primary language $i$ in a country $A$, $d_{AB}$ is Euclidean distance between each pair of counties, $N_{ij}$ is the number of country pairs where $i$ or $j$ are spoken as primary language. &
The larger the language-speaking population and the smaller the distance between the countries $A,B$, the more the attraction between $i$ and $j$. Based on the countries' primary languages.
& Country population data is taken from CIA Factbook \cite{CIA_factbook}.
\\
\midrule
H5: Shared religion    \begin{equation*}  t_{ij}= \begin{cases}    1,           & \text{if }  r_{i}=r_{j}\\    0  & \text{otherwise}\end{cases}\end{equation*}&
$r_i$ is the dominating religion of a language community. It is defined as the most common religion in the list of countries whose primary language is $i$. &
Cultures which profess the same religion will show consistent interest in the same topics.
& The data on world religions was taken from the most recent 2010 Report on Religious Diversity provided by the Pew Research Center \cite{data:religion}.
\\
\bottomrule
\end{tabular}
}}
\end{table}

% Fig 3

%\textbf{Data. }
%To formalise some of the hypotheses, we needed to map languages to countries where they are spoken. Territory--language information was downloaded from \cite{data:lang}, and is based on the data from the World Bank, Ethnologue, FactBook, and other sources, including per-country census data. The data on language family classification was taken from English Wikipedia infoboxes of articles on each of 110 languages, such as 'Hebrew language'. Distance between countries is computed as Euclidean distance in kilometers between country capitals \cite{CIA_factbook}. The data on world religions was taken from the most recent 2010 Report on Religious Diversity provided by the Pew Research Center \cite{data:religion}.

\subsection{Bayesian inference -- HypTrails} \label{Hyptrails}
In order to explain why certain languages form communities of shared interest, we need to explain the link weights, or $z$-score values. We formulate multiple hypotheses based on real-world statistical data, and compare their plausibility using HypTrails \cite{hyptrails}, a Bayesian approach based on Markov chain processes. We input the $z$-scores into a matrix, and express hypotheses about their values via Dirichlet priors -- matrices of transition probabilities between each possible state (in our case -- language edition). We use the trial roulette method to compare different hypothesis. This approach allows to visualise how plausibility of the hypotheses changes with the increasing belief and decreasing allowed variation. Although it was initially designed to compare hypotheses about human trails, in this paper we show that HypTrails is also useful in explaining link weights in networks.

\textbf{Data preparation. }
Using the formalisations detailed in Table \ref{hypotheses_table}, we fill out corresponding transition probabilities matrices. We apply Laplacian smoothing of weight 1 to all matrices to avoid sparsity issues and to account for the cases when editions co-edit a topic of a general encyclopedic importance which might be relevant for multiple language communities. All matrices are normalised row-wise; diagonals are zero as no self-loops are allowed.

%%%%%%%%%%%%%%%%%%%%%%%%
\textbf{Hyptrails ranking. }
The Hyptrails algorithm does not output the absolute values for plausibility of hypotheses, but only compares them one to another. Thus, one must always compare the hypotheses to a uniform hypothesis, and discard those hypotheses that are ranked below the uniform. For the upper bound of comparison, we use the $z$-scores data itself, since no hypothesis can explain the data better than the data itself.

\begin{figure}
  \centering
  \includegraphics[width=0.70\textwidth]{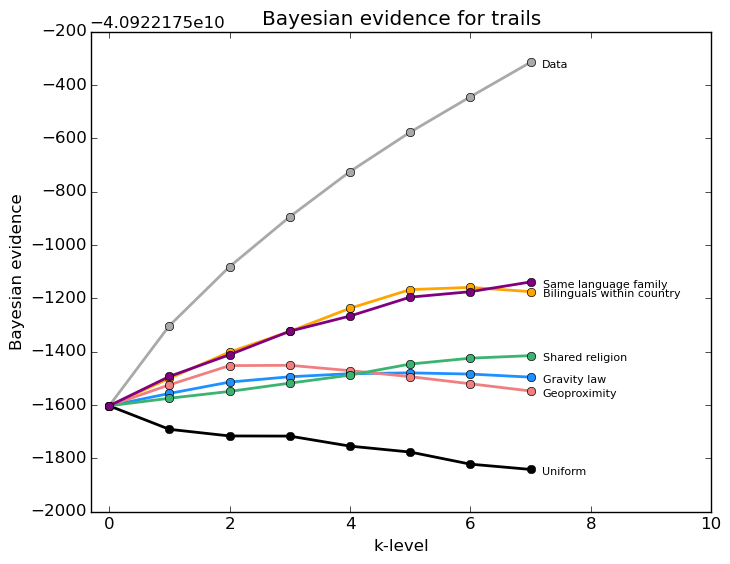}
  \caption{
     \csentence{HypTrails-computed Bayesian evidence for hypotheses plausibility on shared editing interest Wikipedia data.}
Higher values of the Bayesian evidence denote that a hypothesis fits the data well. The bottom black line represents the hypothesis of random shared interests and the top grey line is the fit of data on itself -- together forming an upper and lower limit for fitting hypothesis. The ranking of hypotheses should be compared for the same $k$. All hypotheses are significant, but the most plausible ones to explain cultural proximity are the shared language family, the bilingual, the shared religion, and the gravity law hypotheses. The results show that cultural factors such as language and religion play a larger role in explaining Wikipedia co-editing than geographical factors. }
  \label{fig:hyptrails}
\end{figure}

The results suggest that multiple factors play role in how shared interests are shaped, including geographical proximity, population attraction, shared religion, and especially strongly, linguistic relatedness of the languages and the number of bilingual speakers. No hypothesis explains perfectly all variations in the data, however and all Bayes Factors for all pairs of hypotheses are decisive. Geographical proximity only explains the data to a limited extent, and decays for higher values of $k$, while the number of bilinguals in the same country, shared language family, and shared religion hypotheses grow stronger with more belief, which suggests that they explain the data most robustly. The explanatory power of hypotheses should be compared for the same values of $k$, which expresses how strongly we believe in the hypotheses and how much variation is allowed. Fig.~\ref{fig:hyptrails} summarises the results of the HypTrails algorithm. All hypotheses are compared against the uniform hypotheses of random co-occurrence.

% Fig 4

\subsection{Frequentist approach -- MRQAP} \label{mrqap}
In addition to the HypTrails analysis, we use Multiple Regression Quadratic Assignment Procedure (MRQAP) \cite{hubert1976} to assess statistical significance of association between the concept co-editing network ties and various hypothesis. This method has a long established tradition in social network analysis as a way to sift out spuriously observed correlations~\cite{dekker2003}, and is well-suited for analysing dyadic data where observations are autocorrelated if they are in the same row or column \cite{krackardt1987qap}. We treat the network of concept co-editing as a dependent variable matrix; the independent variable contains the set of hypotheses about the configuration of the network, expressed via hypotheses matrices. Formulation of hypotheses is given in Table~\ref{hypotheses_table}. We normalise the matrices row-wise in order to standardise the values across matrices. MRQAP is a nonparametric test -- it permutes the dependent variables to account for dyadic inter-dependencies. It is also robust against various underlying data distributions \cite{dekker2007sensitivity}. We used 1,000 permutations, which usually suffices for the procedure \cite{jackson1989}.

\textbf{MRQAP ranking. }
The results of the test are in agreement with the hypothesis ranking obtained from applying HypTrails. The number of bilinguals, shared language family, shared religion and demographic attraction are the factors significantly contributing to cultural similarity, as suggested by the $t$-statistic. By including all five hypotheses into Model 1, we are able to explain 15\% of variation in the data. Geographical distance, although a significant factor in several models, is not a very strong one: after excluding the distance hypothesis (Model 2), precision does not decrease. Excluding other hypotheses one by one (Models 3, 4, 5 and 6) lowers precision considerably. Finally, shared language family and bilinguals alone (Models 21 and 22) explain 5\% and 7\% variation in shared interests correspondingly. The results of the MRQAP are reported in Table \ref{table:mrqap}. Different models include variations of hypotheses combinations that explain the variation in language co-editing ties.

% Table 2
\begin{table}
\centering
\caption{MRQAP decomposition of pairwise correspondence between concept co-occurrence and cultural factors. The combination of all hypotheses explains most of the variation in the data (15\%). The most plausible explanations are the number of bilinguals and shared religion. The results of MRQAP agree with the ranking of hypotheses by the HypTrails algorithm. All statistics except those labelled with $^*$ are significant at the 0.05 level.
\label{table:mrqap}
}
\makebox[\textwidth]{
\scalebox{0.90}{
\begin{tabular}{ l l | rrrrr | rrrr }

\toprule
\multicolumn{2}{c|}{\textbf{Model}} &
\textbf{Bilinguals} &
\textbf{Lang. family} &
\textbf{Religion} &
\textbf{Gravity} &
\textbf{Distance}\textsuperscript{1} &
\textbf{$R^2$ adj.} & \textbf{F-stat.} & \textbf{dF} & \textbf{Intercept} \\
\midrule
\midrule
1 & Estimate      & $0.0688$ & $0.1074$ & $0.0900$ & $0.0470$ & $-0.0042^*$ & $\mathbf{0.1458}$ & $410.3$ & $11984$ & $0.0066$ \\
  & $t$-statistic & $\mathbf{27.6524}$ & $\mathbf{23.6158}$ & $\mathbf{13.4772}$ & $\mathbf{10.2732}$ & $\mathbf{-1.3422^*}$ & \\
\midrule
\midrule
2 & Estimate      & $0.0676$ & $0.1075$ & $0.0894$ & $0.0464$ & --$^{\phantom *}$ & $0.1458$ & $512.4$ & $11985$ & $0.0067$ \\
  & $t$-statistic & $29.1517$ & $23.6428$ & $13.4200$ & $10.1893$ & --$^{\phantom *}$ & \\
\midrule
3 & Estimate      & $0.0703$ & $0.1129$ & $0.1022$ & -- & $-0.0009^*$ & $0.1384$ & $482.3$ & $11985$ & $0.0067$ \\
  & $t$-statistic & $28.1932$ & $24.8853$ & $15.4831$ & -- & $-0.2989^*$ & \\
\midrule
4 & Estimate      & $0.0685$ & $0.1080$ & -- & $0.0581$ & $-0.0016^*$ & $0.1329$ & $460.5$ & $11985$ & $0.0074$ \\
  & $t$-statistic & $27.3119$ & $23.5817$ & -- & $12.7773$ & $-0.5225^*$ & \\
\midrule
5 & Estimate      & $0.0716$ & -- & $0.0916$ & $0.0598$ & $-0.0055^*$ & $0.1061$ & $356.9$ & $11985$ & $0.0075$ \\
  & $t$-statistic & $28.1697$ & -- & $13.4180$ & $12.8396$ & $-1.7256^*$ & \\
\midrule
6 & Estimate      & -- & $0.1134$ & $0.0881$ & $0.0546$ & $0.0272^{\phantom *}$ & $0.09140$ & $302.5$ & $11985$ & $0.0070$ \\
  & $t$-statistic & -- & $24.2095$ & $12.7958$ & $11.5815$ & $9.0453^{\phantom *}$ & \\
\midrule
7 & Estimate      & $0.0700$ & $0.1129$ & $0.1020$ & -- & --$^{\phantom *}$ & $0.1386$ & $643.1$ & $11986$ & $0.0067$ \\
  & $t$-statistic & $30.2487$ & $24.8885$ & $15.5098$ & -- & --$^{\phantom *}$ & \\
\midrule
8 & Estimate      & $0.0703$ & $0.1151$ & -- & -- & $0.0030^*$ & $0.1212$ & $552.2$ & $11986$ & $0.0076$ \\
  & $t$-statistic & $27.9237$ & $25.1460$ & -- & -- & $0.9388^*$ & \\
\midrule
9 & Estimate      & -- & -- & $0.0898$ & $0.0684$ & $0.0272^{\phantom *}$ & $0.0470$ & $198.2$ & $11986$ & $0.0079$ \\
  & $t$-statistic & -- & -- & $12.7323$ & $14.2619$ & $8.8191^{\phantom *}$ & \\
\midrule
10 & Estimate      & $0.0700$ & -- & $0.0909$ & $0.0590$ & --$^{\phantom *}$ & $0.1060$ & $474.8$ & $11986$ & $0.0075$ \\
   & $t$-statistic & $29.5521$ & -- & $13.3370$ & $12.7297$ & --$^{\phantom *}$ & \\
\midrule
11 & Estimate      & -- & $0.1140$ & -- & $0.0654$ & $0.0296^{\phantom *}$ & $0.0790$ & $344.0$ & $11986$ & $0.0077$ \\
   & $t$-statistic & -- & $24.1755$ & -- & $13.9808$ & $9.7791^{\phantom *}$ & \\
\midrule
12 & Estimate      & $0.0712$ & $0.1151$ & -- & -- & --$^{\phantom *}$ & $0.1212$ & $827.8$ & $11987$ & $0.0076$ \\
   & $t$-statistic & $30.4703$ & $25.1430$ & -- & -- & --$^{\phantom *}$ & \\
\midrule
13 & Estimate      & $0.0738$ & -- & -- & -- & $0.0027^{\phantom *}$ & $0.0749$ & $486.5$ & $11987$ & $0.0085$ \\
   & $t$-statistic & $28.6184$ & -- & -- & -- & $0.8295^{\phantom *}$ & \\
\midrule
14 & Estimate      & -- & -- & -- & $0.0794$ & $0.0296^{\phantom *}$ & $0.0342$ & $213.4$ & $11987$ & $0.0086$ \\
   & $t$-statistic & -- & -- & -- & $16.7162$ & $9.5508^{\phantom *}$ & \\
\midrule
15 & Estimate      & $0.0733$ & -- & $0.1072$ & -- & --$^{\phantom *}$ & $0.0940$ & $622.8$ & $11987$ & $0.0076$ \\
   & $t$-statistic & $30.9368$ & -- & $15.9020$ & -- & --$^{\phantom *}$ & \\
\midrule
16 & Estimate      & -- & $0.1222$ & -- & -- & $0.0357^{\phantom *}$ & $0.0641$ & $411.6$ & $11987$ & $0.0080$ \\
   & $t$-statistic & -- & $25.9063$ & -- & -- & $11.8512^{\phantom *}$ & \\
\midrule
17 & Estimate      & -- & -- & $0.0936$ & $0.0741$ & --$^{\phantom *}$ & $0.0409$ & $256.8$ & $11987$ & $0.0080$ \\
   & $t$-statistic & -- & -- & $13.2534$ & $15.5280$ & --$^{\phantom *}$ & \\
\midrule
18 & Estimate      & -- & -- & -- & -- & $0.0372^{\phantom *}$ & $0.0118$ & $144.1$ & $11988$ & $0.0090$ \\
   & $t$-statistic & -- & -- & -- & -- & $12.0025^{\phantom *}$ & \\
\midrule
19 & Estimate      & -- & -- & -- & $0.0861$ & --$^{\phantom *}$ & $0.0269$ & $333.1$ & $11988$ & $0.0087$ \\
   & $t$-statistic & -- & -- & -- & $18.2514$ & --$^{\phantom *}$ & \\
\midrule
20 & Estimate      & -- & -- & $0.1144$ & -- & --$^{\phantom *}$ & $0.0217$ & $267.1$ & $11988$ & $0.0081$ \\
   & $t$-statistic & -- & -- & $16.3447$ & -- & --$^{\phantom *}$ & \\
\midrule
21 & Estimate      & -- & $0.1233$ & -- & -- & --$^{\phantom *}$ & $0.0532$ & $674.9$ & $11988$ & $0.0081$ \\
   & $t$-statistic & -- & $25.9798$ & -- & -- & --$^{\phantom *}$ & \\
\midrule
22 & Estimate      & $0.0746$ & -- & -- & -- & --$^{\phantom *}$ & $0.0749$ & $972.2$ & $11988$ & $0.0085$ \\
   & $t$-statistic & $31.1808$ & -- & -- & -- & --$^{\phantom *}$ & \\
\bottomrule
~\\
\multicolumn{10}{l}{\textsuperscript{1} primary language}
\end{tabular}
}}
\end{table}

%%%%%%%%%%%%%%%%%%%%%%%%%%%%%%%%%%%          Discussion          %%%%%%%%%%%%%%%%%%%%%%%%%%%%%%%%%%%%%%%%%%%%%%%%%%%%%%%%%%%%
\section{Discussion} \label{discussion}
In this paper, we have used edit co-occurrences data to investigate cultural similarities between language communities on Wikipedia. We have applied a statistical filtering approach to quantify co-editing similarities and build a network of mutual interests. We have utilised the logic of Bayesian and frequentist hypothesis testing to examine what societal features can explain the observed language clusters. Both approaches render similar results, suggesting that cultural proximity and similarity of interests are best explained by bilingualism, linguistic relatedness of languages, shared religion, and demographic attraction of communities. Geographical distance is a weak, and not very significant factor.

\textbf{Limitations. }
Our study is not free of limitations, some of which are inherent to the nature of the chosen data. Although we found in the literature mounting evidence that Wikipedia is a promising and rich data source for those interested in mining cultural relations, we agree that it is only one of many possible media where culture might find reflection. Moreover, Wikipedia itself is not free from structural biases, as it reflects the activity of selected technology-savvy, mostly white and male \cite{Hill-7, Antin-1}, educated, and economically stable social elites. It by no means is representative of the views of general population. However, it is the elites that often drive the cultural, political, and economic processes \cite{Ronen}, and thus Wikipedia editors represent a group worthy of being studied. Furthermore, we point out that even though we focus on 110 largest language editions, we still compare the editions at different growth stages and levels of topical saturation. Although this might introduce unforeseen biases, we do not see it as a major limitation, since we focus on aggregated editing activity and only on the articles created between 2005 and 2013. We leave for future research the interesting task of incorporating the time dimension in the analysis and examining how interests shape and change over time.

Additionally, while our approach is quantitative, it requires some subjectivity in interpreting the clusters and formulating hypotheses. To strengthen the internal validity of the study, we inform our reasoning about the hypotheses both in visual analysis of the clusters and in previous literature on the subject. Still, we do not claim to have exhausted all possible hypotheses which could explain the data. Moreover, other formalisations of the selected hypotheses might render different results.

One of the benefits of our approach is that it is free of biases related to topic selection, since we avoid focusing on specific kinds of topics where cultural similarities might be expected. It also scales well in terms of the number of communities and hypotheses that could be analysed. In case of research on multilingual data, an important benefit of our approach is that it only uses metadata on user interactions, and understanding the language itself is not required. Finally, it is applicable for any example of collaborative production of a common good where individual activity of participants is recorded.

\textbf{Discussion of results.} Culture is a very complex concept without a definition that is unanimously accepted by Anthropologists, Social Scientists, or Linguists. Although it is universally agreed that cultural communities exist, their borders are very fuzzy and depend on how the researcher defines the term 'culture'. In this work, we focus on the relation between language and culture, and particularly, on how online linguistic expressions can help distil cultural similarities between multilingual communities of Wikipedia editors. An inseparable part of culture, language is only one way of cultural expression, and more studies are needed to explore how other aspects of culture manifest themselves in off- and online world.

Our analysis shows that the decision to write or not to write an article on a certain topic is not a random one. Similar to the idea of national cultural repertoires in the traditional Cultural Sociology \cite{Lamont-10}, we find that various linguistic communities apply different grammars of worth and criteria of evaluation when selecting the topics to cover, that would appeal to the common interest of the language community. Thus, each language edition represents a community of shared understanding with unique linguistic point of view \cite{Bao-2012, Massa-2012, Massa-2011}, its own controversial topics \cite{Yasseri-book}, and concept coverage \cite{Callahan-2011}.

We demonstrate that similarity of co-editing interests between language communities can be partially explained by the number of bilinguals and by linguistic similarity of the languages themselves. This comes as little surprise, since language is a fundamental part of identity, self-recognition, and culture \cite{Kramsch, Castells-2011, Whorf, Bloomfield}. It is hard to separate the effects of the number of bilinguals and shared language family from one another, since both might be related: shared vocabulary and grammatical features of the languages from the same language family might explain higher level of bilingualism for these language dyads. Moreover, language choice and bilingualism are an effect of factors galore, such as post-colonial history, education, language and human right policies, free travel, and migration due to political instability, poverty, religious persecutions or work \cite{Rassool-1998, Crystal-2000}. Finally, cultural similarity defined through Hofstede's four dimensions of values \cite{Hofstede-1980} has also been found to relate to language~\cite{Pfeil-2006, West-2004}.

Shared religion is another uniting factor for language communities. Our finding is in line with Huntington's thesis which argues that cultural and religious identities of people form the primary source of potential conflict in the post-Cold War era~\cite{Huntington}. The studies of email and Twitter communication \cite{State-2015} and similarity in country information interests \cite{Karimi} also reveal the patterns that echo religious ``fault lines''.

Population attraction and geographical proximity are the uniting factors that have been extensively discussed in the literature, most relevantly in the context of mobile communication flows \cite{Krings-2009} and migration \cite{Simini-2012}. Similar to our results, several studies report gravity laws in online settings, including \cite{Backstrom} and  \cite{Karimi}. Not only choice of topics to edit, but also online trade in taste-dependent products is affected by distance. For example, \cite{Blum-2006} finds that proximate countries show more similarity in taste. Notably, this effect only holds for culture-related products such as music. This further supports our finding that there is a relationship between geographical distance and culture, and allows us to speculate that the Internet fails to defy the law of gravity.

The question of whether English is becoming the world's \textit{lingua franca} is an intriguing one \cite{crystal:global}. Its central, influential position in the global language network has been reported in networks of book translations, multilingual Twitter users, and Wikipedia editors \cite{Ronen, Hale-2014, Hale-2014-Twitter}. On the one hand, such high visibility allows information to radiate between the more connected languages. On the other hand, our study shows that global language centrality plays a minor role in shared interests. Moreover, we show that the domination of English disappears in the network of co-editing similarities, and instead local interconnections come to the forefront, rooting in shared language, similar linguistic characteristics, religion, and demographic proximity. A similar effect has been observed in international markets, where economic competitiveness is linked to the ability to speak a local \textit{lingua franca}, rather than English \cite{Bel-2011}.

\section{Conclusions and Implications} \label{conclusions}
Out of almost 300 Wikipedia's language editions, 76\% have less than 100 active users \cite{WP:list}. Linguistically, this means that those languages are in danger of extinction \cite{Crystal-2000}, at least in the online space \cite{Netlang}. Nevertheless, \cite{Kornai-2013} emphasises the role of Wikipedia in helping peripheral languages cross the digital divide, acquire digital functions and prestige as their speakers go online. At the same time, Pentzold \cite{Pentzold-12} describes Wikipedia as a global cultural memory place, access to which depends on the language skills. In his view, Wikipedia is not a mere encyclopedia where facts are documented, but rather a space where the entire collective memories of important events are constructed during a discursive, social process. We show that the topics that each language edition documents are not selected randomly, however small the underlying community of editors. These non-random processes might relate to the fact that each Wikipedia language edition presents a cultural memory place, where the linguistic point of view and the memorable events of that community are negotiated.

Our findings bring some important policy questions for the Wikimedia Foundation, such as: What are the cultural implications of populating editions with automatically translated concepts present in other language editions? Should English Wikipedia aim at becoming an all-inclusive collection of information from other language editions? Should the decision on who and what will be remembered belong to the community of editors, however small, or to an automated algorithm? We hope that our research will inspire dialogue on how similarities  between language communities can be used to improve participation of editors speaking peripheral languages and expand the content of smaller editions.

In addition, Wikipedia has a power to mobilise cultural communities around a very important collective task -- selecting and archiving important knowledge for future generations. Our analysis sheds light on how cultural similarities are reflected in this process. We also demonstrate that global cultural interconnections are not dominated by one powerful player, but instead follow the locally established ``fault lines'' of bilingualism, shared religion and population attraction. We hope that these results will be useful for managers, economists and politicians working in multicultural settings, enthusiastic Wikipedians, academics wishing to study culture via the web, as well as for the public curious about global, intercultural relationships.

%%%%%%%%%%%%%%%%%%%%%%%%%%%%%%%%%%%%%%%%%%%%%%
%%                                          %%
%% Backmatter begins here                   %%
%%                                          %%
%%%%%%%%%%%%%%%%%%%%%%%%%%%%%%%%%%%%%%%%%%%%%%

\begin{backmatter}
\section*{Competing interests}
The authors declare that they have no competing interests.

\section*{Author's contributions}
AS, MS, FK conceived and designed the research. AS acquired the data. AS, FK and DE analysed the data. AS and FK interpreted the results. All authors discussed, wrote, and approved the final version of the manuscript.

\section*{Acknowledgements}
We would like to thank Michael Macy, Florian Lemmerich, and Philipp Singer for inspiring discussions and useful comments. We also thank the Wikimedia Foundation for developing and hosting the Wikimedia Labs infrastructure and granting access to its servers. JK acknowledges the funding from the European Community's Seventh Framework Programme under grant agreement n\textsuperscript{o}~610928, REVEAL.

%%%%%%%%%%%%%%%%%%%%%%%%%%%%%%%%%%%%%%%%%%%%%%%%%%%%%%%%%%%%%
%%                  The Bibliography                       %%
%%                                                         %%
%%  Bmc_mathpys.bst  will be used to                       %%
%%  create a .BBL file for submission.                     %%
%%  After submission of the .TEX file,                     %%
%%  you will be prompted to submit your .BBL file.         %%
%%                                                         %%
%%                                                         %%
%%  Note that the displayed Bibliography will not          %%
%%  necessarily be rendered by Latex exactly as specified  %%
%%  in the online Instructions for Authors.                %%
%%                                                         %%
%%%%%%%%%%%%%%%%%%%%%%%%%%%%%%%%%%%%%%%%%%%%%%%%%%%%%%%%%%%%%

%%%%%%%%%%%%%%%%%%%%%%%%%%%%%%%%%%%    Bibliography  %%%%%%%%%%%%%%%%%%%%%%%%%%%%%%%%%%%%%%%%%%%%%
\bibliographystyle{bmc-mathphys} % bmc-mathphys
\bibliography{Samoilenko_submission}      % bibliography_WP_cross-cultural_analysis_paper

%%%%%%%%%%%%%%%%%%%%%%%%%%%%%%%%%%%
%%                               %%
%% Additional Files              %%
%%                               %%
%%%%%%%%%%%%%%%%%%%%%%%%%%%%%%%%%%%
\clearpage
\section*{Appendix}
%% \textbf{Additional file 1 (.pdf)} -- Figure A1.
%% The figure A1 contains the heatmap of editing co-occurrence comparison between empirical and experimental data based on a 6.5\% sample of the data (N = 200,748 concepts).

%% \textbf{Additional file 2 (.pdf)}. -- Table A1.
%% The table A1 contains the clusters of languages with shared interest as found by the Infomap clustering algorithm.

\setcounter{figure}{0}
\renewcommand{\thefigure}{A\arabic{figure}}%
\begin{figure}[h]
  \centering
  \includegraphics[width=.98\textwidth]{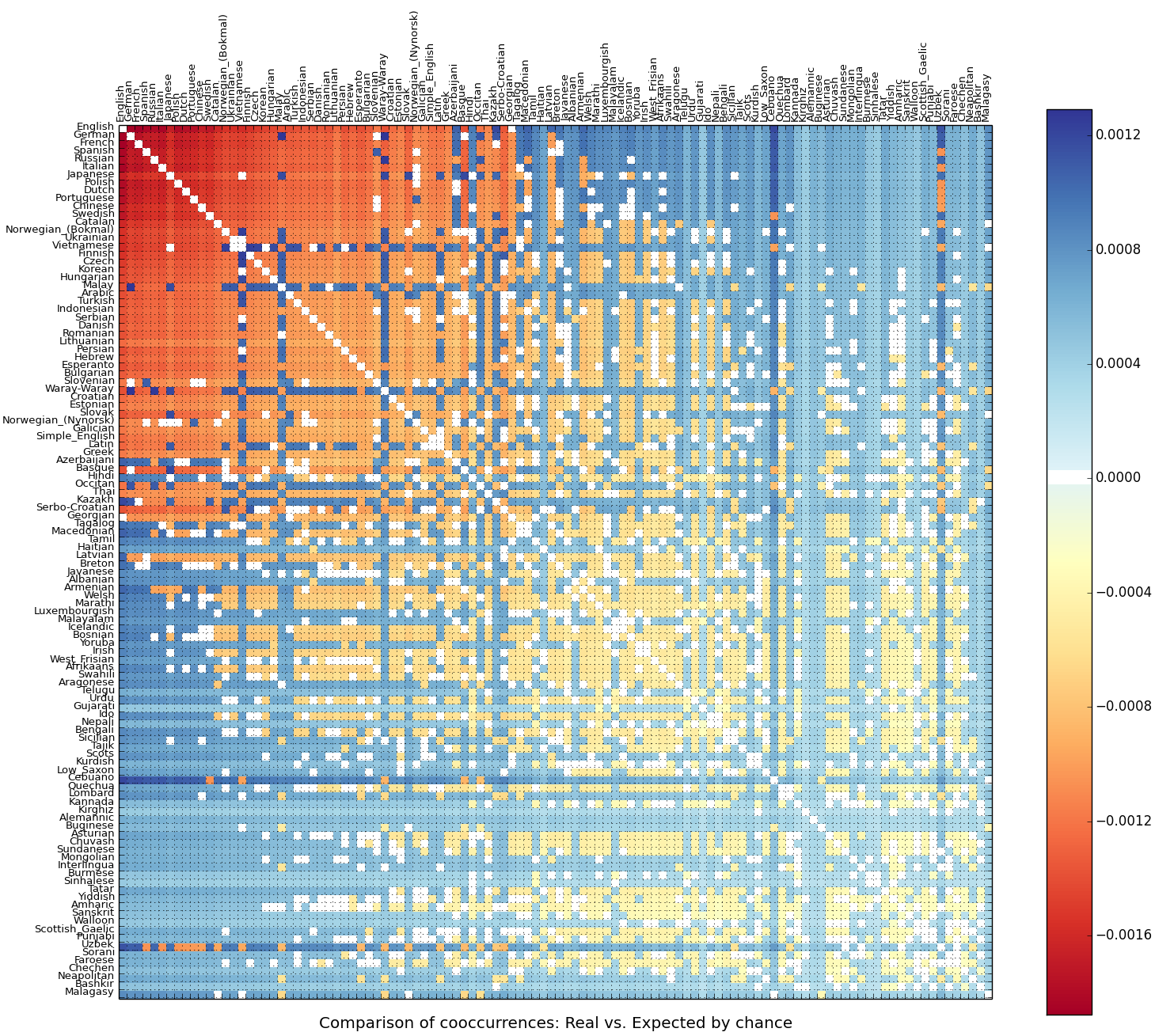}
  \caption{\csentence{Comparison of empirical and experimental data on editing co-occurrences.} White cells are explained by the null model, shades of blue/red show the distance of observed co-occurrences from the lower/upper border of the confidence interval. Low explained variation (11\%, 95\% confidence level), suggests that non-random processes are in place. Based on a random 6.5\% data (N = 200,748 concepts).}
  \label{fig:SI}
\end{figure}

\setcounter{table}{0}
\renewcommand{\thetable}{A\arabic{table}}%
\begin{table}[h]
\centering
\caption{Clusters of languages with shared interest as found by the Infomap clustering algorithm. The weight of each language is the normalized weighted degree of the node. Some languages, including English, do not belong to a larger community and form a self-cluster instead. }
\label{SI_table}
\scalebox{0.91}{
\begin{tabular}{c@{\hspace{1cm}}c}
\begin{minipage}[t][][b]{6cm}
\begin{tabular}{rll}
\toprule
\textbf{Cluster} & \textbf{Language}& \textbf{Weight}  \\
\midrule
\midrule
1 & French & 0.01415 \\
& Occitan & 0.01372 \\
& Waray-Waray & 0.01291 \\
& Latin & 0.01219 \\
& Italian & 0.01147 \\
& Cebuano & 0.00652 \\
& Alemannic & 0.00591 \\
& Tagalog & 0.00581 \\
& Lombard & 0.00566 \\
& Sicilian & 0.00536 \\
& Buginese & 0.00394 \\
& Neapolitan & 0.00355 \\
\midrule
2 & Russian & 0.01665 \\
& Polish & 0.01584 \\
& Ukrainian & 0.01509 \\
& Armenian & 0.01190 \\
& Estonian & 0.01068 \\
& Lithuanian & 0.00994 \\
& Latvian & 0.00800 \\
\midrule
3 & Sanskrit & 0.01046 \\
& Hindi & 0.01032 \\
& Tamil & 0.00927 \\
& Malayalam & 0.00869 \\
& Telugu & 0.00861 \\
& Marathi & 0.00826 \\
& Bengali & 0.00613 \\
& Kannada & 0.00578 \\
& Nepali & 0.00578 \\
& Gujarati & 0.00537 \\
& Punjabi & 0.00528 \\
& Sinhalese & 0.00337 \\
\midrule
4 & Norwegian (Bokmal) & 0.01818 \\
& Swedish & 0.01499 \\
& Norwegian (Nynorsk) & 0.01311 \\
& Finnish & 0.01093 \\
& Danish & 0.01048 \\
& Icelandic & 0.00640 \\
& Faroese & 0.00432 \\
\midrule
5 & Serbian & 0.01884 \\
& Serbo-Croatian & 0.01632 \\
& Croatian & 0.01417 \\
& Slovenian & 0.01133 \\
& Bosnian & 0.01036 \\
& Haitian & 0.00568 \\
\midrule
6 & Vietnamese & 0.01362 \\
& Turkish & 0.01022 \\
& Persian & 0.00966 \\
& Azerbaijani & 0.00965 \\
& Arabic & 0.00886 \\
& Urdu & 0.00838 \\
& Kurdish & 0.00629 \\
& Tajik & 0.00523 \\
& Sorani & 0.00423 \\
\midrule
7 & Esperanto & 0.01717 \\
& Hungarian & 0.01552 \\
& Czech & 0.01361 \\
& Slovak & 0.01159 \\
& Romanian & 0.01104 \\
\bottomrule
\end{tabular}
\end{minipage}
&
\begin{minipage}[t][][b]{6cm}
\begin{tabular}{rll}
\toprule
\textbf{Cluster} & \textbf{Language} & \textbf{Weight}    \\
\midrule
8 & Catalan & 0.01566 \\
& Galician & 0.01011 \\
& Basque & 0.00983 \\
& Spanish & 0.00903 \\
& Aragonese & 0.00864 \\
& Asturian & 0.00576 \\
\midrule
9 & Chechen & 0.01332 \\
& Bashkir & 0.01191 \\
& Tatar & 0.01184 \\
& Kazakh & 0.01005 \\
& Chuvash & 0.00878 \\
\midrule
10 & Indonesian & 0.02004 \\
& Malay & 0.01410 \\
& Javanese & 0.01310 \\
& Sundanese & 0.00773 \\
\midrule
11 & Bulgarian & 0.01232 \\
& Macedonian & 0.01135 \\
& Greek & 0.00878 \\
& Albanian & 0.00688 \\
\midrule
12 & Chinese & 0.00891 \\
& Japanese & 0.00792 \\
& Korean & 0.00734 \\
& Thai & 0.00617 \\
\midrule
13 & Breton & 0.00948 \\
& Welsh & 0.00648 \\
& Scottish Gaelic & 0.00509 \\
& Irish & 0.00494 \\
\midrule
14 & Dutch & 0.01673 \\
& West Frisian & 0.00630 \\
\midrule
15 & German & 0.01354 \\
& Low Saxon & 0.00520 \\
\midrule
16 & Georgian & 0.01042 \\
& Quechua & 0.00451 \\
\midrule
17 & Uzbek & 0.00797 \\
& Kirghiz & 0.00385 \\
\midrule
18 & Hebrew & 0.00847 \\
& Yiddish & 0.00298 \\
\midrule
19 & Luxembourgish & 0.00710 \\
& Walloon & 0.00295 \\
\midrule
20 & Ido & 0.00612 \\
& Interlingua & 0.00390 \\
\midrule
21 & Afrikaans & 0.00740 \\
& Amharic & 0.00229 \\
\midrule
22 & Portuguese & 0.00892 \\
\midrule
23 & Simple English & 0.00806 \\
\midrule
24 & English & 0.00763 \\
\midrule
25 & Swahili & 0.00560 \\
\midrule
26 & Scots & 0.00486 \\
\midrule
27 & Yoruba & 0.00453 \\
\midrule
28 & Mongolian & 0.00349 \\
\midrule
29 & Burmese & 0.00238 \\
\midrule
30 & Malagasy & 0.00187 \\
\bottomrule
\end{tabular}
\end{minipage}
\end{tabular}
}
\end{table}

\end{backmatter}
\end{document}